# SNAP Focal Plane

M. Lampton[f], C. Bebek[a], C. Akerlof[b], G. Aldering[a], R. Amanullah[c], P. Astier[d], E. Barrelet[d], L. Bergström[c], J. Bercovitz[a], G. Bernstein[e], M. Bester[f], A. Bonissent[g], C. Bower[h], W. Carithers[a], E. Commins[f], C. Day[a], S. Deustua[i], R. DiGennaro[a], A. Ealet[g], R. Ellis[j], M. Eriksson[c], A. Fruchter[k], J-F. Genat[d], G. Goldhaber[f], A. Goobar[c], D. Groom[a], S. Harris[f], P. Harvey[f], H. Heetderks[f], S. Holland[a], D. Huterer[l], A. Karcher[a], A. Kim[a], W. Kolbe[a], B. Krieger[a], R. Lafever[a], J. Lamoureux[f], M. Levi[a], D. Levin[b], E. Linder[a], S. Loken[a], R. Malina[m], R. Massey[n], T. McKay[b], S. McKee[b], R. Miquel[a], E. Mörtsell[c], N. Mostek[h], S. Mufson[h], J. Musser[h], P. Nugent[a], H. Oluseyi[a], R. Pain[d], N. Palaio[a], D. Pankow[f], S. Perlmutter[a], R. Pratt[f], E. Prieto[m], A. Refregier[n], J. Rhodes[o], K. Robinson[a], N. Roe[a], M. Sholl[f], M. Schubnell[b], G. Smadja[p], G. Smoot[f], A. Spadafora[a], G. Tarle[b], A. Tomasch[b], H. von der Lippe[a], D. Vincent[d], J-P. Walder[a], G. Wang[a]

[a] Lawrence Berkeley National Laboratory, Berkeley CA, USA
[b] University of Michigan, Ann Arbor MI, USA
[c] University of Stockholm, Stockholm, Sweden
[d] CNRS/IN2P3/LPNHE, Paris, France
[e] University of Pennsylvania, Philadelphia PA, USA
[f] University of California, Berkeley CA, USA
[g] CNRS/IN2P3/CPPM, Marseille, France
[h] Indiana University, Bloomington IN, USA
[i] American Astronomical Society, Washington DC, USA
[j] California Institute of Technology, Pasadena CA, USA
[k] Space Telescope Science Institute, Baltimore MD, USA
[l] Case Western Reserve University, Cleveland OH, USA
[m] CNRS/INSU/LAM, Marseille, France
[n] Cambridge University, Cambridge, UK
[o] NASA Goddard Space Flight Center, Greenbelt MD, USA
[p] CNRS/IN2P3/INPL, Lyon, France

**ABSTRACT**

The proposed SuperNova/Acceleration Probe (SNAP) mission will have a two-meter class telescope delivering diffraction-limited images to an instrumented 0.7 square-degree field sensitive in the visible and near-infrared wavelength regime. We describe the requirements for the instrument suite and the evolution of the focal plane design to the present concept in which all the instrumentation – visible and near-infrared imagers, spectrograph, and star guiders – share one common focal plane.

**Keywords**: focal plane detectors, CCD, HgCdTe

# 1. INTRODUCTION

The SuperNova/Acceleration Probe (SNAP) is a planned space flight mission intended to obtain a precise measurement of the expansion of the Universe. The experiment is prompted by the discovery[1,2] that this expansion is evidently accelerating, not decelerating as would be predicted by models that contain mass-energy but no cosmological constant or dark energy. To effectively test models of the expansion, it is essential to compare accurate observational data against model predictions of the expansion rate as a function of lookback time, or equivalently, luminosity distance. Type Ia supernovae (SNe) populate the observable Universe and serve as accurately standardizable candles. Each measured SN furnishes a redshift and a peak magnitude. The redshift is a measure of the expansion between the SN epoch and the present, while the magnitude is a measure of the distance to theSN, and hence the elapsed time since the SN exploded. Properly calibrated and sorted into systematic classes, a collection of a few thousand such SNe spanning the redshift range $0.3<z<1.7$ will provide important new constraints on models of the universe and the dark energy that it contains. A more detailed description of the mission and its science is presented elsewhere in these proceedings[3-7]; see also the SNAP web page at http://snap.lbl.gov.

The requirements placed on the SNAP instrument derive directly from the science goals. The primary requirement is to obtain the peak brightness vs. redshift of at least 2000 Type Ia SNe out to a redshift of $z=1.7$. Identification of Type Ia SNe requires the measurement of characteristic features in their spectra, taken near peak luminosity. Host-galaxy redshift is also determined spectroscopically. The peak brightness is experimentally taken to be the flux through a B-band filter defined in the rest-frame of the SN. For SNe at $z>1$, the rest-frame B-band is red-shifted to the near-infrared. Photometry utilizing a range of filters is necessary to synthesize a common B-band rest-frame magnitude over the range $0.3 < z < 1.7$; this procedure is known as "K-correction."[8] A fit is then performed to a series of K-corrected magnitudes taken over the rise and fall of the supernova lightcurve, in order to extract the peak magnitude with an accuracy of ~2%. Spectra of a few supernovae will be taken at several points over the light curve to provide spectral templates to improve the accuracy of the K-corrections. A correction for host-galaxy extinction is performed by comparing the flux in the B-band to the flux in other, redder bands; the shorter wavelength B-band light is preferentially scattered by galactic dust. From this final, fitted and corrected peak magnitude and the measured host-galaxy redshift, a Hubble diagram of time vs. expansion rate is constructed and can be fit to various cosmological models to constrain the nature of the dark energy.

# 2. THE SNAP INSTRUMENT REQUIREMENTS

To meet the scientific requirements imposed by the program of measurements described above, SNAP has a large, 0.7 square degree instrumented field of view and an observation cadence of 4 days, commensurate with SN evolution times. Discovery and photometric follow-up are automatically accomplished with the large field-of-view imager that repetitively scans a fixed region of sky. A spectrometer optimized for SN spectra is allocated observation time for follow-up spectroscopy and template building. The derived requirements for the SNAP instrument described here are preliminary and subject to change during the conceptual design phase.

Figure 1 shows critical points on the light curve and the desired measurement accuracy. We note that the stated signal-to-noise ratio (S/N) need not be achieved with a single measurement but can be synthesized from multiple measurements, taking advantage of the substantial time dilation for high-redshift SNe. A more detailed discussion of the importance of the measurements taken at each epoch can be found in a companion paper.[3]

The SNAP imager addresses the above requirements using two detector technologies to efficiently cover the wavelength range of 350 nm to 1700 nm. The NIR range (900 nm to 1700 nm) is measured with HgCdTe devices.[4] The HgCdTe devices have a fixed pixel size of 18 µm and the telescope optics are designed to give an angular pixel size of 0.17 arcsec, a good match to the telescope diffraction limit at 1700 nm.[5] The visible region (350 nm to 1000 nm) is measured with CCDs designed at Lawrence Berkeley National Laboratory.[9] We have control of the CCD pixel size and have chosen 10.5 µm so as to have a good match to the telescope diffraction limit at 1000 nm with an angular pixel size of 0.1 arcsec.

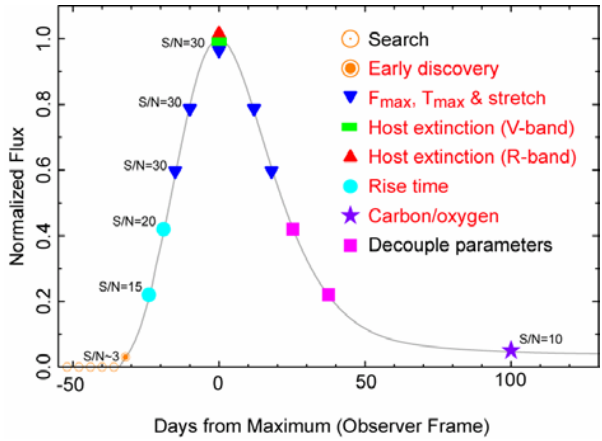 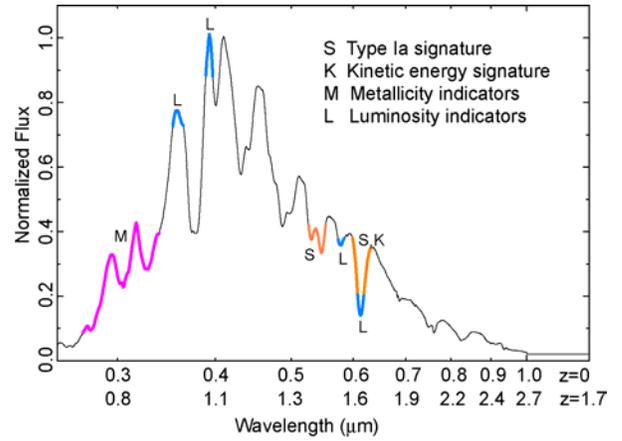

**Figure 1.** Sample B-band light curve for a z=0.8 Type Ia supernova. S/N targets at different epochs are shown.

**Figure 2.** Type Ia spectra showing critical features. The horizontal axis shows wavelengths for z=0 and z=1.7

To achieve accurate K-corrections, nine logarithmically-distributed B-band filters have been found to be sufficient. The S/N versus SN epoch is achieved by having sufficiently long exposure times and repeating them at regular four-day intervals. Imager detector specifications are given in Table 1.

**Table 1.** Mission reference specifications for the imager and its sensors.

| Parameter | Visible | NIR | Units |
|---|---|---|---|
| Field of View | 0.34 | 0.34 | $deg^2$ |
| Plate scale (nominal) | 0.10 | 0.17 | arcsec |
| Wavelength | 350-1000 | 900-1700 | nm |
| <Quantum efficiency> | 80 | 60 | % |
| Read noise (multiple reads) | 4 | 5 | e |
| Dark current | 0.002 | 0.02 | e/s/pixel |
| Filters | 6 | 3 | |

The spectrometer is used to make a positive identification of Type Ia SNe by observing the characteristic SiII feature at 6150 Å. Other features in the spectrum of a typical Type Ia SNe, shown in Figure 2, will allow a more detailed characterization of parameters such as metallicity and ejecta velocities. The restframe UV side of the spectrum extends down to 350 nm for z=0.3 while the signature SiII line is at 1700 nm for z=1.7. This wavelength range is covered by a two-channel spectrometer.[6] Because the UV side of the SN light curve is always within the wavelength range of the SNAP CCDs, these features benefit from the greater quantum efficiency of the CCD relative to a HgCdTe sensor, significantly reducing the required exposure time to achieve the desired S/N ratio. The CCD channel of the spectrograph covers 350 nm to 980 nm while a HgCdTe channel covers 980 to 1700 nm; a very modest resolution ($\lambda/\delta\lambda$) of 100 is sufficient. An integral field unit is used, implemented as an image slicer, to reduce the absolute pointing requirements of the satellite and to allow simultaneous measurement of the host galaxy spectra. Table 2 shows the specifications for the spectrograph.

**Table 2.** Mission reference specifications for the spectrograph and its sensors.

| Parameter | Visible | NIR | Units |
|---|---|---|---|
| Wavelength coverage | 350-980 | 980-1700 | nm |
| Plate scale | 0.15 | 0.15 | arcsec |
| Spatial resolution | 0.15 | 0.15 | arcsec |
| Field-of-View | 3 x 3 | 3 x 3 | arcsec$^2$ |
| Resolution | 100 | 100 | $\lambda/\delta\lambda$ |
| <Quantum Efficiency> | 80 | 60 | % |
| Read Noise | 2 | 5 | e |
| Dark Current | 0.001 | 0.02 | e/s/pixel |

## 3. FOCAL PLANE CONCEPT DEVELOPMENT

The present SNAP focal plane arrangement has evolved over the last two years. The same set of instruments – visible and NIR imagers, spectrograph, and star guiders – has been present in all the concepts. An immutable aspect of the instrument operations is repetitive imaging of fixed regions of the sky. Transients identified as Type Ia supernovae by ground-based analysis of the imaging data are targeted into a spectrograph near peak brightness. The repetitive imaging simultaneously accomplishes discovery and follow-up of supernovae using multiple filters.

The starting concept of the SNAP instruments had separate focal planes for the different detectors; the various components tapped the telescope light at a multitude of points. We had several concerns with this configuration:

- Multiple focal planes requiring simultaneous focus and pointing stability.
- Multiple filter and shutter mechanisms.
- Non-overlapping fields of view for the visible and NIR imagers.
- Small field of view for the NIR imager, consisting only of one to four sensors.
- Large observational inefficiency from targeting SNe onto the NIR imager.

One characteristic of this concept was that it could be pointed anywhere in the sky to produce a complete set of multi-filter measurements in one or the other imager. Rotation of the satellite around the primary axis did not affect the observation plan.

To address the concerns above, we studied alternative concepts in which all imager sensors, the spectrograph input, and the star guiders are mounted at a common focal plane. The following describes different strategies that we considered for deployment of sensors, filters, and shutters around such an integrated focal plane. The distinctions between these strategies mainly concern the imagers and how to filter and shutter their light. The spectrograph imposes no preference for one solution over another.

One strategy involved different mixtures of visible and NIR sensors behind a filter wheel and shutter. We identified several difficulties with this strategy:

- Different fields of view for visible and NIR.
- Time-consuming cross-measurement of stars at visible and NIR wavelengths, further complicated by the inefficient geometry of the annular focal plane.
- Large filter wheel with filters ~200 mm in diameter. Without multi-bandpass filters, one sensor type or the other would be blind at any given time.

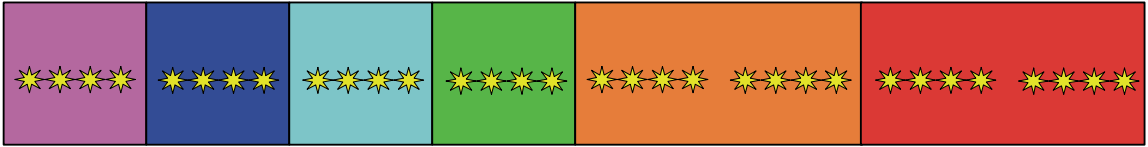

**Figure 3.** A simple example depicting a linear array of fixed filters. Shown are the positions of a single star as the filters are dragged past it.

To deal with the latter issue, we considered a second strategy with filter wheels and wedge-shaped filters located above the focal plane. An outer annulus contained CCDs and the inner annulus contained HgCdTe NIR sensors. The filter wheels were also divided into an inner and outer annulus, with filter areas weighted to achieve the required S/N in each bandpass (more red filter area to accommodate the lower photon flux in the z-shifted B-band spectrum of high z SNe). A motivation for this design was to address actuator failure. If a filter wheel stuck, steering the satellite to move stars across the filters could partially save the mission. The visible and NIR fields of view, however, were still disjoint, and the first two difficulties identified for the initial strategy would still apply.

We did consider the option of using a single sensor maintaining sensitivity across the full wavelength span of interest, 350 nm to 1700 nm. This would allow all sensors to remain active at any filter setting and allow the overlap of visible and NIR fields of view. However, the broad wavelength span coupled with a fixed pixel size dramatically impacted the SNAP mission. For efficient pixel use we have found that sampling the Airy diffraction disk at full-width zero max with one pixel is sufficient (*i.e.*, a factor of two undersampling). We apply this at 1000 nm for the visible detectors and at 1700 nm for the NIR. A single pixel size at our present NIR 0.17 arcsec angular scale would require twice as many exposures (dithering) to generate the same photometric accuracy in the blue. The complementary approach would match the 0.1 arcsec/pixel scale of the visible imager. This would require a trebling of the focal plane area to cover the same field of view, or a 65% reduction of the field of view
to maintain the same focal plane area.

All the single-focal-plane strategies described so far involve large, massive, untried filter wheels. We and external advisors viewed this as a high risk. Also, over the last year, the science team has concluded that each SNe needs to be measured with a larger number of filters in order to build light curves in multiple colors (*e.g.*, dust extinction correction uses differences between multiple pairs of filters). The blue end of the SNe spectra, restframe U-band, has grown in importance for classification as well as restframe R- and I-band. At high z, the population of non-Type Ia SNe is expected to be large, and false identification is expensive in time consumed for spectrographic measurement of high z objects. Measurements in the restframe R-band appear to provide a powerful handle on dust extinction determination. We reached the consensus that ideally all objects should be measured in all filters, visible and NIR. The above led us to consider a class of focal planes with fixed filters.

## 4. FIXED-FILTER FOCAL PLANE

A simple example of a fixed-filter focal plane is shown in Figure 3. A linear array of sensors with different filters is swept across a star field by steering the telescope. The actual motion is a series of steps, each step having a fixed exposure time. The number of steps within a filter determines the total integration time. Note that multiple steps within a filter facilitate standard techniques for optimizing photometric accuracy such as dithering. For example, four exposures within a filter could be offset precisely from each other by a fraction of a pixel size. A coarse hop to the next filter then would follow.

In this simple example, some filters are longer than others. This is a way to gain longer integration time in the redder NIR filters, where SN signals are weaker. Here we also point out an inefficiency in any fixed filter scheme: stars are not measured in all filters both at the beginning and at the end of the scan. This inefficiency can be minimized, but not eliminated, by scanning a long, linear piece of the sky.

To populate a useful focal plane, a two- rather than a one-dimensional deployment of filters is required. It should be obvious that no array of square filters can repetitively measure the same patch of sky in all filters if the satellite rotates continuously relative to the observation fields. It is not a constraint, however, that the satellite rotate continuously. Even though SNAP has fixed, body mounted solar cells and a passive heat radiator, the satellite orientation relative to normal solar incidence can tolerate rotations up to ±45°. The satellite and focal plane can be held in fixed alignment relative to the observation field for a three month period, after which the satellite is rotated 90°. Figure 4 clarifies this point.

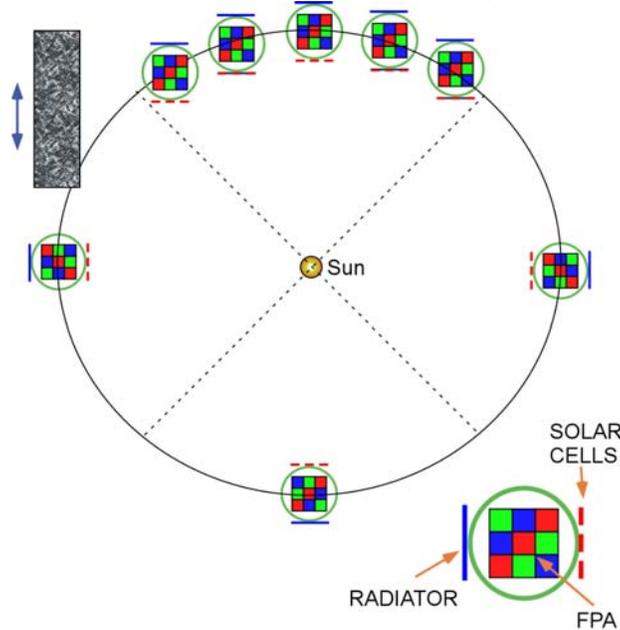

**Figure 4.** An illustration of the interaction between the satellite and its focal plane array, the orientation to the Sun, and an observation field (upper left). The upper arc shows how the satellite must rotate relative to the Sun to maintain the fixed filters in a constant orientation relative to the observation field. The left and right arcs show a 90° rotation to maintain the general orientation of the solar cells towards the Sun. The double arrow next to the observation field shows the direction in which the satellite is rocked to scan the field.

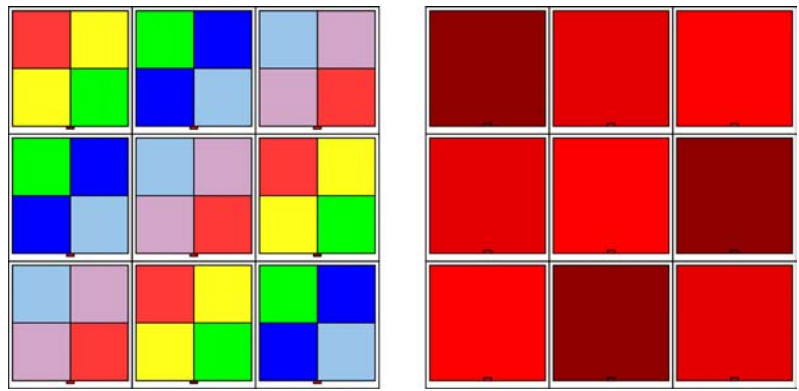

**Figure 5.** On the left is a two axis symmetric deployment of six filters types for the visible imager such that vertical or horizontal scans of the array through an observation field will measure all objects in all filters. On the right is the same concept for an array of three filter types for the NIR imager. The false colors indicate filters with the same bandpass.

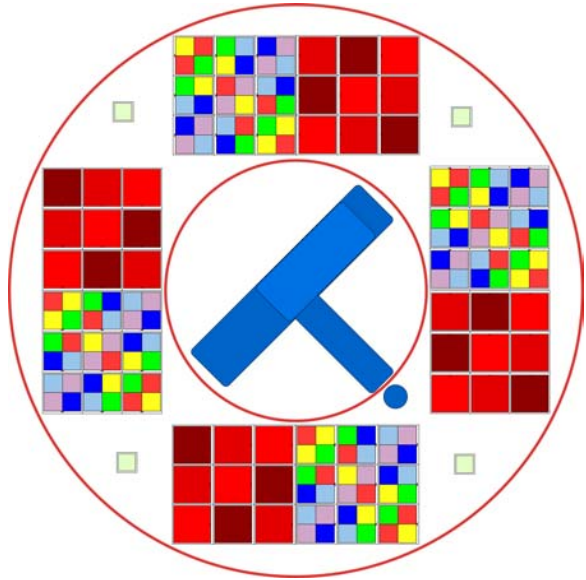

**Figure 6.** The SNAP focal plane working concept. The two axis symmetry of the imager filters allows any 90° rotation to scan a fixed strip of the sky and measure all objects in all nine filter types. Underlying the filters, there are 36 2k x 2k HgCdTe NIR devices and 36 3.5k x 3.5k CCDs colocated on a 140K cooled focal plane. The central rectangle and solid circle are the spectrograph body and its light access port, respectively. The spectrum of a supernova is taken by placing the star in the spectrograph port by steering the satellite. The four small, isolated squares are the star guider CCDs. The inner and outer radii are 129 and 284 mm, respectively.

The constraint that the satellite be rotated be in 90° increments requires the filter pattern to remain symmetric with respect to two orthogonal axes. Consider either of the filter arrays in Figure 5. Note that the arrays can be scanned though an observation field left-to-right, right-to-left, top-to-bottom, and bottom-to-top, and that a given star will be measured with each filter bandpass but not necessarily the same physical filter. Note that any 90° rotation of the filter array can still measure the star field in all filter types.

A study was performed to determine the minimum filter set required. This is primarily determined by the precision needed for the K-correction, the reconstruction of the restframe B-band light from a set of laboratory-frame filter measurements. The study found that six visible filters and three NIR filters are sufficient if they are derived from a B-band filter with logarithmic (1+z) scaling of their wavelength centers and widths. Figure 6 shows an array of visible and NIR filters. To enhance the amount of NIR light that is integrated, the individual NIR filters have twice the area of the individual visible filters and hence integrate twice as many exposures.

The HgCdTe sensor physical size is fixed by the vendor. We have control of the CCD size, so that it is possible to make the physical dimensions of the 6x6 CCD filter array equal to those of the 3x3 NIR filter array. Adopting this choice, there are only two low order solutions for populating our annular focal plane with these filter unit cells. Figure 6 shows the solution that makes most efficient use of the available field of view (the rejected solution is half as efficient). Underlying each NIR filter is one 2k x 2k, 18 μm HgCdTe device, 36 in total. Underlying each 2x2 array of visible filters is one 3.5k × 3.5k, 10.5 μm CCDs, also 36 in total. The total number of pixels is ~600 million.

The working concept for the SNAP focal plane as shown in Figure 6 would be passively cooled to operate at 140K Short flex cables penetrate the focal plane to bring the signals to the electronics located on the backside. The present conceptual design envisions an ASIC-based readout that would also operate at 140K; this avoids routing low-level analog signals long distances and reduces the size of the cable plant between the cold focal plane and the warm data acquisition electronics located to the side.

## 5. SIMULATED PERFORMANCE

We have established that the concept in Figure 6 can meet the photometric S/N goals for SNAP as shown in Figure 1 for SNe with z<1.7. Inputs to this finding are the imager detector specifications in Table 1, the nine filters arrived at to implement K-corrections, and an observational strategy using four, 300 s exposures for each filter in the visible, corresponding to eight, 300 s exposures in each NIR filter (with twice the filter area). The SNAP instrument is in a high earth orbit that allows it to be in observing mode 86% of the time; 60% of the observing time is dedicated to repetitive photometric imaging and 40% is spent doing targeted spectrographic follow-up. Assuming a SN explosion rate that is constant in redshift and co-moving volume,[10] Figure 7 shows the SNAP SN discovery rate per year for one particular observation scenario. A SNAP reference mission spends 16 months observing a north ecliptic field and 16 months observing a south ecliptic field resulting in a total of 5800 SNe discovered and 4200 with follow-up spectroscopy.

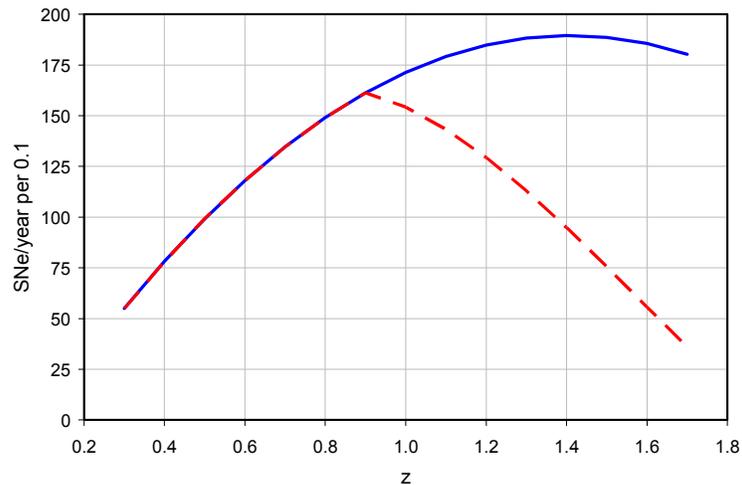

**Figure 7.** The number of supernovae discovered and followed (solid curve) per 0.1 redshift bin and the number with spectroscopic information (dashed curve) assuming a 60%-40% split of total observation time between photometry and spectroscopy. 2200 SNe are discovered in 7.5 sq deg and 1600 have spectroscopic follow-up in a one year period.

## ACKNOWLEDGMENTS

This work was supported by the Director, Office of Science, of the U.S. Department of Energy under Contract No. DE-AC03-76SF00098.

## REFERENCES

1. S. Perlmutter *et al*., Astrophys.J., 517, p.565, 1999.
2. A. G. Riess *et al*., Astron.J., 116, p.1009. 1998.
3. G. Aldering *et al*., Proc. SPIE **4835**-21, 2002.
4. G. Tarle *et al.*, Proc. SPIE **4850**-131, 2002.
5. M. Lampton *et al*., Proc. SPIE **4849**-29, 2002.
6. A. Ealet *et al.*, Proc. SPIE **4850**-165, 2002.
7. A. Kim *et al.,* Proc. SPIE **4836**-10, 2002.
8. P. Nugent, A. Kim, S. Perlmutter, PASP **114**:803 (2002)
   A. Kim, A. Goobar and S. Perlmutter, PASP **108**: 190 (1996).
9. D. Groom, *et al.,* NIM **A442**, p. 16, 2000.
10. R. Pain, *et al.,* astro-ph/0205476 (accepted for publication in *The Astrophysical Journal,*).